\documentclass[iop,referee,onecolumn]{emulateapj}
\usepackage{natbib}
\usepackage{ulem,footnote,lineno,graphicx,color,rotating,amssymb,amsmath,amsfonts,times,hyperref}
\newcommand{\lsim}{\mathrel{\mathop{\kern 0pt \rlap
  {\raise.2ex\hbox{$<$}}}
  \lower.9ex\hbox{\kern-.190em $\sim$}}}
\newcommand{\gsim}{\mathrel{\mathop{\kern 0pt \rlap
  {\raise.2ex\hbox{$>$}}}
  \lower.9ex\hbox{\kern-.190em $\sim$}}}

\shorttitle{Diffuse $\gamma$-ray emission from unresolved BL Lac objects}
\shortauthors{Di Mauro et al.}

\begin{document}

\title{Diffuse $\gamma$-ray emission from unresolved BL Lac objects}
\author{M. Di Mauro
$^{1,2}$, F. Donato$^{1}$, G. Lamanna$^{3}$, D. A. Sanchez$^{3}$, P. D. Serpico$^{2}$}
\affil{$^1$Physics Department, Torino University, and  Istituto Nazionale di Fisica Nucleare, Sezione di Torino, via Giuria 1, 10125 Torino, Italy\\
$^2$Laboratoire d' Annecy-le-Vieux de Physique Th\'eorique (LAPTh), Univ. de Savoie, CNRS, B.P.110, Annecy-le-Vieux F-74941, France\\
$^3$Laboratoire d' Annecy-le-Vieux de Physique des Particules (LAPP), Univ. de Savoie, CNRS/IN2P3, Annecy-le-Vieux F-74941, France}

\preprint[LAPTH-067/13

\email{mattia.dimauro@to.infn.it (corresponding author)}

  \begin{abstract}
Blazars, active galactic nuclei with a jet pointing toward the Earth, represent the most abundant class of high-energy
extragalactic $\gamma$-ray sources. The subset of blazars known as BL Lac objects is on average closer to Earth (i.e. younger) and 
characterized by harder spectra at high energy than the whole sample. The fraction of BL Lacs that is too dim to be detected and resolved by current $\gamma$-ray telescopes is therefore
expected to contribute to the high-energy isotropic diffuse $\gamma$-ray background (IGRB).
The IGRB has been recently measured over a wide energy range by the Large Area
Telescope (LAT) on board the Gamma-ray Space Telescope ({\it
Fermi}). We present a new prediction of the diffuse $\gamma$-ray flux due to the
unresolved BL Lac blazar population. The model is built upon the spectral energy
distribution and the luminosity function derived from the fraction of
BL Lacs detected (and spectrally characterized) in the $\gamma$-ray energy range. We
focus our attention on the ${\cal O}(100)$ GeV energy range, predicting the
emission up to the TeV scale and taking into account the absorption on the
extragalactic background light. In order to better shape the BL Lac spectral energy distribution, we combine the {\it Fermi}-LAT data with Imaging Atmospheric Cerenkov Telescopes
measurements of the most energetic sources. Our analysis is carried on
separately for low- and intermediate-synchrotron-peaked BL Lacs on one hand, and high-synchrotron-peaked BL Lacs on the other one: we find
in fact statistically different features for the two. The diffuse emission
from the sum of both BL Lac classes increases from about 10$\%$ of the measured IGRB
at 100 MeV to $\sim$100$\%$ of the data level at 100 GeV. At energies greater than
100 GeV, our predictions naturally explain the IGRB data, accommodating their
softening with increasing energy. Uncertainties are estimated to be within of a
factor of two of the best-fit flux up to 500 GeV. 
  \end{abstract}
  

\maketitle

\section{Introduction}
\label{sec:intro}
The origin of the observed isotropic diffuse $\gamma$-ray background (IGRB) at
MeV-GeV energies is one of the most intriguing problems in astrophysics. The
presence of an isotropic component was first suggested by the OSO-3 satellite
\citep{1972ApJ...177..341K} and then confirmed by SAS-2
\citep{1975ApJ...198..163F} and EGRET \citep{1998ApJ...494..523S}. The {\it
Fermi}-Large Area Telescope (LAT) has provided a measurement of this isotropic component showing that
it can be adequately described as a single power law with an index of $-2.41\pm
0.05$ in the 200 MeV-100 GeV energy range \citep{2010PhRvL.104j1101A}. Recently
the {\it Fermi}-LAT Collaboration has presented a new estimation of the IGRB,
based on 44 months of data in the range 200 MeV-400 GeV: at energy $E\geq100$ GeV, the data
points fall systematically below the extrapolation of the low-energy powerlaw best-fit
 \citep{ackermann2012} \footnote{\url{http://galprop.stanford.edu/download/talks/13.Fermi_symposium_Ackermann.pdf}}.
We underline that these data are preliminary and error bands include systematics
from effective area uncertainty and cosmic-ray background subtraction but not from foreground model uncertainties.

In several models, a significant fraction of the IGRB is attributed to
unresolved extragalactic $\gamma$-ray sources. Blazars, belonging to the large
family of active galactic nuclei (AGNs), represent the most numerous identified
source population in the EGRET \citep{1999ApJS..123...79H} and {\it Fermi}-LAT
\citep{2012ApJS..199...31N} catalogs, and are expected to produce a substantial
fraction of the IGRB. Typical predictions for blazars contributions to the IGRB
range from $20\%$ to $30\%$
\citep{1998ApJ...496..752C,2000MNRAS.312..177M,narumoto2006,2007ApJ...659..958D,Kneiske:2007jq,2011PhRvD..84j3007A,2009ApJ...702..523I}
to $100\%$ \citep{1996ApJ...464..600S,2011ApJ...736...40S,Neronov:2011kg}. Recently, an
analysis of the source-count distribution at flux
levels\footnote{$F_{100}$ represents the flux integrated above a
threshold energy of 100 MeV.} $F_{100} \geq 10^{-9}$ photons cm$^{-2}$ s$^{-1}$
estimated that the contribution of unresolved blazars to the IGRB is $\sim16\%$
in the 100 MeV-100 GeV band~\citep{2010ApJ...720..435A}. Since the source-count
distributions show a strong break at $F_{100} \geq 6\times 10^{-8}$ photons
cm$^{-2}$ s$^{-1}$, it was concluded that, extrapolating the source-counts to
zero flux, the maximum contribution would be $\sim23\%$ of the IGRB.

An accurate modeling of the blazar contribution to the IGRB requires a careful
consideration of their phenomenological classification. The spectral energy
distributions (SEDs) of blazars exhibit a bimodal structure with a low-energy
component attributed to synchrotron emission, while the nature of the high-energy emission is still under debate \citep{1992AA...256L..27D,1993ApJ...404L..41H,1990ApJ...362...38B,2003AA...410..397S}. Blazars are traditionally divided into
flat-spectrum radio quasars (FSRQs) and BL Lacertae (BL Lac) objects according
to the presence or absence of strong broad emission lines in their optical/UV
spectrum, respectively \citep{1980ARA&A..18..321A,1995PASP..107..803U}. Extending a scheme proposed
for BL Lacs \citep{1995ApJ...444..567P}, all the blazars may as well be
classified according to the value of the synchrotron-peak frequency $\nu_S$,
defining low-synchrotron-peaked (LSP) blazars when the peak frequency is
observed in the far-infrared (FIR) or IR band ($\nu_S < 10^{14}$ Hz),
intermediate-synchrotron-peaked (ISP) blazars when $\nu_S$ is in the
near-IR to ultraviolet (UV) frequencies ($10^{14} \,\,\rm{Hz}\leq \nu_S <
10^{15}$ Hz), or as high-synchrotron-peaked (HSP) blazars if the peak frequency
is located in the UV band or higher energies ($\nu_S \geq 10^{15}$
Hz)~\citep{2010ApJ...716...30A}. 
Almost all the FSRQs with an SED classification are classified as LSP \citep{2011ApJ...743..171A}.

Recently, \citet{2012ApJ...751..108A} examined the properties of
$\gamma$-ray-selected FSRQs using data from the {\it Fermi}-LAT instrument. This
work relies on a sample of 186 FSRQs detected by {\it Fermi}-LAT at high
significance and large Galactic latitude during the first year of operation
\citep{2010ApJS..188..405A} and has thus important consequences for our
understanding of the origins of the IGRB. The SEDs of all FSRQs show some
curvature, with a peak in the 10 MeV-10 GeV range followed by a decrease leading
to undetectable fluxes at energies higher than $\sim 30$ GeV. Moreover, the
study of the FSRQ luminosity function (LF) shows a redshift distribution peaked
at a redshift $z\sim1$, with the farthest sources found at redshift values
$\sim3.1$ and $\gamma$-ray luminosity $\sim10^{50}$ erg s$^{-1}$
\citep{2011ApJ...743..171A,2012ApJ...751..108A}. The modeled SED and LF lead to
a predicted contribution of the FSRQs to the IGRB of $\sim10\%$ in the 1 MeV-10
GeV band. This analysis is in good agreement with the results reported by
\citet{2010ApJ...720..435A}, except above 10 GeV where the use of a simple power law for the FSRQ spectra proved inadequate. Due to their redshift
distribution and the absorption of $\gamma$-rays by the extragalactic background
light, FSRQs are expected to give a negligible contribution to the IGRB above 10 GeV.

BL Lac blazars show different features if compared with FSRQs, and they are thus
expected to give a different contribution to the IGRB. Their redshift
distribution peaks around $z \sim 0.2$ and extends up to $z \approx 1.5$, at significantly lower redshifts than FSRQs
\citep{2011ApJ...743..171A}. Also, marked differences exist in the distributions of  $\gamma$-ray
luminosity $L_{\gamma}$ and of spectral index $\Gamma$. The
$L_{\gamma}$ distributions range between $L_{\gamma}\in[10^{45},10^{49}]$ erg s$^{-1}$
and $L_{\gamma}\in[10^{44},10^{47}]$ erg s$^{-1}$ for FSRQs and BL Lacs, respectively,
while the corresponding $\Gamma$ distributions peak at $2.42\pm0.17$ for FSRQs
and it splits into $2.17\pm0.12$, $2.13\pm0.14$, and $1.90\pm0.17$ for LSP, ISP
and HSP, respectively \citep{2011ApJ...743..171A}. An immediate consequence of both
the distance distribution (nearer for BL Lacs) and the spectral characteristics
(harder for BL Lacs) is that the BL Lac contribution to the IGRB should be
dominant above tens of GeV with respect to the FSRQ one. This is the region we
concentrate on in the rest of this paper, even if we will provide our results on
the diffuse emission from tens of MeV up to tens of TeV. 

The IGRB well beyond tens of GeV (and up to sub-TeV energies) has been paid
relatively minor attention in the past for a number of reasons. Most notably, it
was a difficult range to access experimentally. This situation is changing
thanks to the extended dynamical range and higher statistics of the {\it
Fermi}-LAT instrument on one hand, and to the improved performances of Imaging
Atmospheric Cerenkov Telescopes (IACTs) on the other hand. Although IACTs are
not particularly suitable for diffuse studies, the gradual lowering of energy
thresholds and the increase of the collecting area (thanks to wide-dish
telescopes) is allowing a more thorough exploration of the window of energy
around ${\cal O}(100)$\, GeV. Also, the greatly improved number of sources
discovered in the TeV range allows us to meaningfully combine these data with
{\it Fermi}-LAT results, leading to a more complete spectral characterization
in the energy range of interest.

The aim of the paper is to predict the diffuse $\gamma$-ray emission from
unresolved BL Lacs, in light of the currently available information. We
consider a complete sample of BL Lacs detected during the first two years of
{\it Fermi}-LAT operation to determine the LF and we use {\it Fermi}-LAT
data together with TeV data from Cerenkov telescope, when available, to
determine the SED. To our knowledge, this is the first time that such a synergic
approach has been attempted. The article is structured as follows. Section \ref{LumFunct}
is devoted to the LF determination; Section \ref{lumfun} introduces
the catalogs and the criteria used to select our sample and
Section \ref{lumfunFit} describes our fits to several models. Section \ref{SedSec}
deals with the parameterization of the SEDs: similarly to the structure of the previous section, in
subsection~\ref{SedCat} we describe the data sets used, while
subsection~\ref{SedFit} details our fitting formulae and the results.
Section~\ref{diffuse} illustrates our prediction of the diffuse $\gamma$-ray
emission from unresolved BL Lacs. Finally, Section \ref{concl} includes a
discussion of the (astro)physical relevance of our findings, also comparing our
results with diffuse $\gamma$-ray contributions from other unresolved sources,
and presents some perspectives and conclusions. Some technical results
concerning the fitting formulae of the LF and our HSP catalog
are reported in Appendices A and B, respectively.
\medskip

\section{The $\gamma$-ray Luminosity Function}
\label{LumFunct}

\subsection{The Catalog and Selection Cuts}\label{lumfun}
\begin{figure}
\centering
\includegraphics[width=0.50\textwidth]{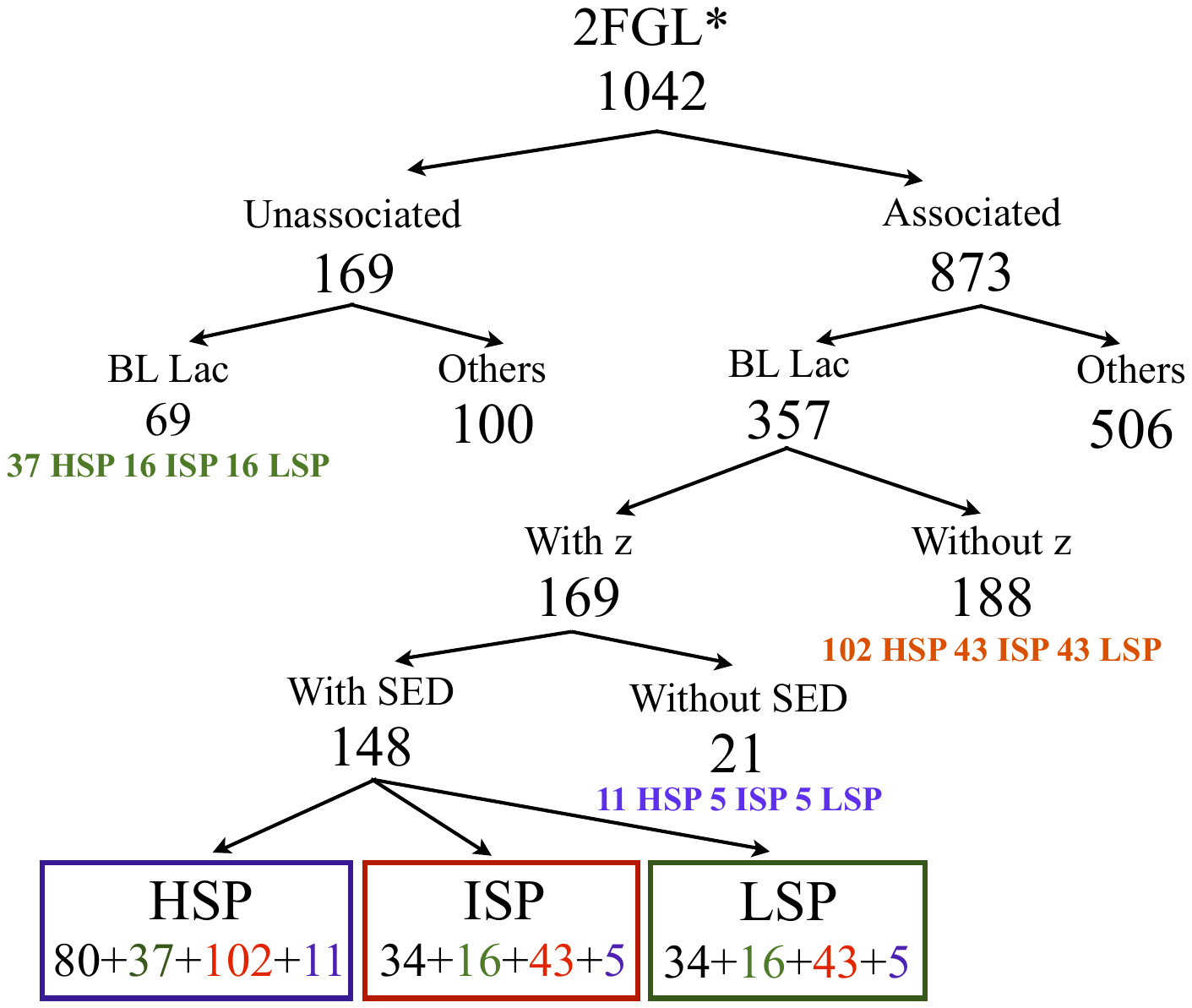}
\caption{In this figure we sketch the source selection criteria we consider in our analysis for the calculation of the $\gamma$-ray LF.
		We list the steps bringing us from our initial catalog 2FGL* (namely, BL Lacs from 2FGL with $|b| > 10^{\circ}$ and TS$\,>25$)
		 to our final sample of sources; these are
		the independent requests of dealing with associated sources, with sources with measured redshift, and with sources with an SED classification. 
		For each step we report also, with different colors, the number of BL Lacs that are discarded and have to be considered to infer the 
		incompleteness factor of the sample (see text for details).}
\label{fig:selection}
\medskip
\end{figure}

For the determination of the $\gamma$-ray LF, we need the source distributions
with respect to the redshift, the $\gamma$-ray luminosity and an SED
classification. We found that all the needed information can be obtained in the
2-year Fermi-LAT Point Source Catalog~\citep[2FGL,][]{2012ApJS..199...31N}, supplemented by two other
catalogs, \citep{2012ApJ...748...49S} for the determination of the redshift
 and the Second Fermi AGN Catalog (2LAC) for the SED classification \citep{2011ApJ...743..171A} of the selected sources.
As a whole, the 2FGL catalog contains 1873
sources, but for all the studies performed here we restrict ourselves to its
subset of reliably detected, high-latitude objects, requirements that
translate numerically into a galactic latitude $|b| > 10^{\circ}$, and value of
the test statistics TS$\,>25$ (corresponding to a significance larger than
$\approx \,5\,\sigma$ \citep{2012ApJS..199...31N}). These cuts select 1042 sources that effectively
constitute our initial catalog, dubbed 2FGL*. We impose further selection
criteria on this data set in order to determine the LF, as represented in
Figure \ref{fig:selection} and described below: in fact,  the
object classification, the distance distribution, and the $\gamma$-ray spectrum
determination are all necessary ingredients in order to use the sample  for our purposes.

First, we note that 169 out of the 1042 sources are unassociated. Among the
associated sources, 357 are BL Lacs (and 318 FSRQs). 
Note that only 43\% of the BL Lac objects have a measured redshift.
Further, only 278 BL Lacs (78\% of the total) have an SED classification (i.e., 79 sources cannot be
classified owing to the lack of archival data), with HSPs representing the largest
subclass (138 sources, 50\% of SED-classified sources), ISPs the second largest
(72 sources, 26\%), and LSPs the smallest (68 sources, 24\%). On the
contrary, FSRQs with SED classification are 72\% of the total and are
essentially all LSPs (99\%). Eventually, we end up with a catalog of 148 BL Lacs, divided
into 34 LSP, 34 ISP and 80 HSP objects. 

The above-mentioned cut factors are, however, important in determining the
incompleteness of the BL Lac sample, which has to be corrected for this effect
when inferring the properties of the whole population. The incompleteness of our
sample is given by three independent factors: \begin{itemize} \item {\it
Unassociated sources.} The 2FGL* catalog contains 169 unassociated sources among
the total of 1042 sources. This allows us to estimate the presence of further
$169\times\,357/(1042-169)\approx69$ BL Lacs among the unassociated sources, and
in particular of about $69\times80/148\approx37$ HPSs, and
$69\times34/148\approx16$ IPSs as well as LPSs. 
The inferred fraction of BL Lac objects in the unassociated sample is thus $\simeq 41\%$, which is compatible with the results derived in~\citep{2013arXiv1306.6529D}.
The relevant incompleteness
factor is $(37+80)/80\approx1.46$ for HSP and $(16+34)/34\approx1.47$ for LSP
and ISP objects. \item {\it Lack of redshift.} For a total sample of 357
associated BL Lacs only 169 have a measured redshift. Among the 188 sources
without a redshift assignment, about 102 are HSP, and 43 are LSP as well as ISP
BL Lacs. The lack of a measured value for the redshift gives an incompleteness
factor of 2.28 for HSPs, and 2.26 for LSPs and ISPs. \item {\it Lack of SED.} We
have considered only the BL Lacs with an SED classification. For a total sample
of 169 BL Lacs with a measured redshift, only 148 have an SED classification.
Among the 21 sources without an SED classification, about 11 are HSPs and 5 are
LSPs and ISPs. The incompleteness factor associated to the lack of an SED
classification is 1.14 for HSP, and 1.15 for LSP and ISP BL Lacs. \end{itemize}
The total incompleteness of the three subclasses can be thus estimated as
$(80+11+102+37)/80\approx2.88$ for HSP and $(34+5+43+16)/34\approx2.88$ for LSP
and ISP. We take into account these factors in the standard way
\citep{2012ApJ...751..108A}, namely, by correcting upward the normalization of
the LF to reflect the inferred actual number of BL Lacs (associated or not,
with or without a measured redshift and an SED classification). Note that due to
the relatively small number of sources on which these correction factors have
been estimated, a Poissonian uncertainty of ${\cal O}$(10\%-20\%) is typically
associated with the average incompleteness factors reported above.

\subsection{Fitting Models for the $\gamma$-Ray Luminosity Function}\label{lumfunFit}

The characteristic parameters of an extragalactic source are the redshift $z$,
the $\gamma$-ray luminosity $L_{\gamma}$ and the parameters describing the SED, notably the 
photon spectral index $\Gamma$. In the following section we demonstrate that the SED is better reproduced by a power law with an exponential cutoff,
so that at least one more parameter is needed. However, for the sake of notational simplicity, we shall omit an explicit mention of the cutoff energy parameter in
the formulae in the remaining of this section. Additionally, in the energetic range 0.1-100 GeV that is used for the derivation of the LF, the $\gamma$-ray luminosity $L_{\gamma}$ and fluxes 
$F_{100}$ are not sensitive to the cutoff, and the spectral index $\Gamma$ actually constitutes the most important parameter. 
The space density of the BL Lac population may be expressed as
\begin{equation}
\label{eq:spden} \Theta_{\gamma}(z,\Gamma,L_{\gamma}) =
\frac{d^3N}{dz\,d\Gamma\,dL_{\gamma}}= \frac{d^2N}{dV\,dL_{\gamma}}
\frac{dV}{dz} \frac{dN}{d\Gamma} = \rho_{\gamma}(z,L_{\gamma}) \frac{dV}{dz}
\frac{dN}{d\Gamma},
 \end{equation} 
where $dV/dz$ is the co-moving volume element
per unit redshift and unit solid angle \citep{1999astro.ph..5116H}, and
$\rho(z,L_{\gamma})$ is the LF defined as the number of sources per unit of
comoving volume $V$ and luminosity $L_{\gamma}$\citep{2001MNRAS.322..536W}:
\begin{eqnarray} \label{eq:lf} \rho(z,L_{\gamma})=\frac{d^2N}{dV\,dL_{\gamma}}
\, . \end{eqnarray} The function $dN/d\Gamma$ is the (intrinsic) photon index
distribution, assumed to be independent of the redshift, and, in analogy with
\citet{2010ApJ...720..435A,2012ApJ...751..108A,2012ApJ...749..151Z}, it is
modeled as a Gaussian: \begin{equation} \label{eq:dNdGamma} \frac{dN}{d\Gamma} =
e^{\frac{(\Gamma - \mu)^2}{2\sigma^2}}\,, \end{equation} where $\mu$ and
$\sigma$ are the mean and the dispersion values of the Gaussian distribution,
respectively, and will be fixed in the following to the best-fit values of the
SED given in Table \ref{tab:valori}.
%
%
%
%

The aim of this section is to determine the space density of BL Lac in Eq.
\ref{eq:spden} as a function of the rest-frame energy 0.1-100 GeV luminosity
($L_{\gamma}$), the redshift ($z$), and the photon index ($\Gamma$), by
comparing different predictions with the observed distributions of BL Lacs with
respect to $L_{\gamma}$, $z$ as well as with the distribution of
the number of sources with a flux higher than $F_{\gamma}$, $N(>F_{\gamma})$. 
In order to describe the form of the LF in Eq. \ref{eq:lf} we considered the
pure luminosity evolution
\citep[PLE,][]{2005AA...441..417H,2003ApJ...598..886U}, the luminosity-dependent
density evolution \citep[LDDE,][]{2003ApJ...598..886U} and the steep-spectrum
radio sources \citep[SSRS,][]{2001MNRAS.322..536W} models. In the PLE model the
redshift evolution is entirely in luminosity, while in the LDDE model the
redshift evolution depends on the luminosity. These two models have been used in
\cite{2012ApJ...751..108A} to derive the $\gamma$-ray diffuse emission from
FSRQs. On the other hand, the SSRS model has been deduced for steep-spectrum
radio sources and has been employed by
\citet{2011ApJ...733...66I,DiMauro:2013xta} to find the diffuse $\gamma$-ray
emission from misaligned AGNs. Further details and formulae are reported in the
Appendix.
 
In order to find which of the three models is a better fit to our sample of BL
Lacs space density, we first build the experimental $dN/dz$, $dN/dL_{\gamma}$
and $N(>F_{\gamma})$ distributions from the following relations
\citep{2010ApJ...720..435A,Ajello:2013lka}:
\begin{eqnarray}
\label{eq:zdistr_exp}
   \displaystyle 	\frac{dN}{dz} = \frac{1}{\Delta\,z} \frac{N_z}{\Delta \Omega},
\end{eqnarray}

\begin{eqnarray}
\label{eq:Ldistr_exp}
   \displaystyle 	\frac{dN}{dL_{\gamma}} = \frac{1}{\Delta\,L_{\gamma}} \frac{N_{L_{\gamma}}}{\Delta \Omega},
\end{eqnarray}

\begin{eqnarray}
\label{eq:ndistr_exp}
   \displaystyle 	N(>F_{\gamma}) = \sum^{N_{F_{\gamma}}}_{i=1} \frac{1}{\Delta \Omega\, \omega(F_{\gamma,i})},
\end{eqnarray}
where $\Delta\,z$ and $\Delta\,L_{\gamma}$ are the redshift and luminosity
widths of the bin, $N_z$ and $N_{L_{\gamma}}$ are the numbers of BL Lacs counted
in that bin of redshift or luminosity and $\Delta \Omega$ is the solid angle
($|b|>10^{\circ}$ in our case). 
The sum of Eq.~\ref{eq:ndistr_exp} is made for all the $i$th source with a flux $F_{\gamma,i}>F_{\gamma}$.
$\omega(F_{\gamma,i})$ is the {\it Fermi}-LAT
efficiency at flux $F_{\gamma,i}$ and $N_{F_{\gamma}}$ is the number of the
sources with a flux higher than $F_{\gamma}$. We use here the estimation of
the efficiency $\omega(F_{\gamma})$ derived in \citet{DiMauro:2013xta}.
For the above-cited experimental distributions we have taken into account the uncertainties associated with the finite number of sources in each bin $\propto \sqrt{N}$ (Poissonian uncertainties) and the uncertainty on the efficiency as in \citep{2012ApJ...751..108A,Ajello:2013lka}.

From Eq.~\ref{eq:spden}, the redshift, luminosity and source-count distributions
can be computed as \citep{2012ApJ...751..108A,Ajello:2013lka}: 
\begin{eqnarray}
\label{eq:zdistr}
   \displaystyle 	\frac{dN}{dz} = \int^{\Gamma_{\rm max}}_{\Gamma_{\rm min}} d\Gamma \int^{L^{\rm max}_{\gamma}}_{L^{\rm min}_{\gamma}} dL_{\gamma} \,
				\Theta_{\gamma}(z,\Gamma,L_{\gamma}) \, \omega(F_{\gamma}),
\end{eqnarray}

\begin{eqnarray}
\label{eq:Ldistr}
   \displaystyle 	\frac{dN}{dL_{\gamma}} = \int^{\Gamma_{\rm max}}_{\Gamma_{\rm min}} d\Gamma \int^{z_{\rm max}}_{z_{\rm min}} dz \,
				\Theta_{\gamma}(z,\Gamma,L_{\gamma}) \, \omega(F_{\gamma}),
\end{eqnarray}

\begin{eqnarray}
\label{eq:ndistr}
   \displaystyle 	N(>F_{\gamma}) = \int^{\Gamma_{\rm max}}_{\Gamma_{\rm min}} d\Gamma \int^{z_{\rm max}}_{z_{\rm min}} dz  \int^{L^{\rm max}_{\gamma}}_{L_{\gamma}(F_{\gamma},z,\Gamma)} dL_{\gamma} \,
				\Theta_{\gamma}(z,\Gamma,L_{\gamma}),
\end{eqnarray}
where the limits of integration are $L^{\rm min}_{\gamma} = 10^{38}$ erg
s$^{-1}$, $L^{\rm max}_{\gamma} = 10^{52}$ erg s$^{-1}$, $z_{\rm min} = 0$,
$z_{\rm max} = 6$, $\Gamma_{\rm min} = 1.2$, and $\Gamma_{\rm max} = 3.0$ (we
have checked that the results of the analysis have a negligible dependence on
the limits of integration). $L_{\gamma}(F_{\gamma},z,\Gamma)$ is the
$\gamma$-ray luminosity for a source with a flux $F_{\gamma}$ in the range 100
MeV-100 GeV, a redshift $z$ and photon index $\Gamma$. The data on the redshift
and $L_\gamma$ are less precise than the ones on the source-count distribution.
On the other side, $dN/dz$ and $dN/dL_{\gamma}$ data are independent from each
other, differently from the $N$ count data, which is a cumulative measure whose
single data depends on all the ones at higher $F_{\gamma}$. 

The fitting procedure consists of varying the values of the LF parameters and
deducing, with the use of the MINUIT minimization package, the best-fit
configuration with the relevant 1$\sigma$ errors fitting the experimental number
counts for the redshift, luminosity and source distributions. We find that the
reduced $\chi^2$ values are 0.49 for the LDDE model (39 degrees of freedom
(d.o.f.)), 1.12 for the PLE model (41 d.o.f.) and 1.74 for the SSRS model (39 d.o.f.).
The LDDE model turns out to be the best LF in reproducing the whole BL Lac
population (similar results have been obtained very recently in
\cite{Ajello:2013lka}). As will discussed in detail in Sect.
\ref{SedFit}, it is convenient to merge the LSP and ISP BL Lac classes into a
unique population which we will define LISP. We have therefore also fitted the
different LF models separately for the two sub-classes represented by the LISP
and HSP BL Lacs. The same results found for the whole sample hold true. The
LDDE model is the preferred one, with reduced $\tilde{\chi}^2$ of 0.37 (32
d.o.f.) and 0.34 (30 d.o.f.) for LISP and HSP sources, respectively.

\begin{figure}
\centering
\includegraphics[width=0.63\textwidth]{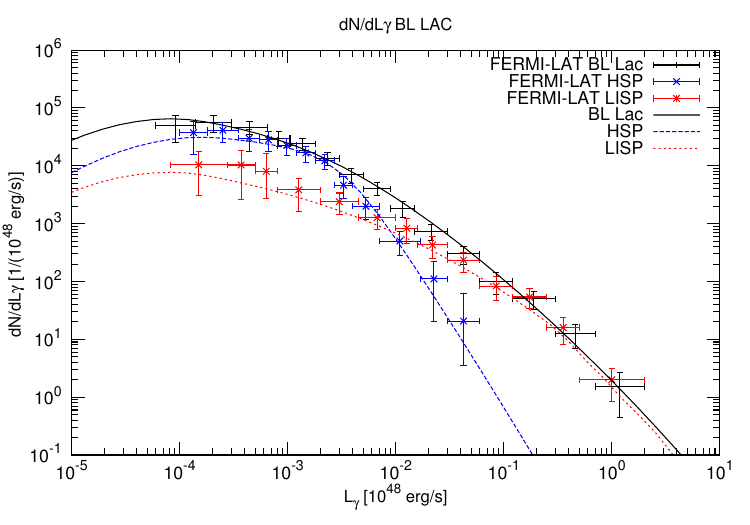}
\includegraphics[width=0.63\textwidth]{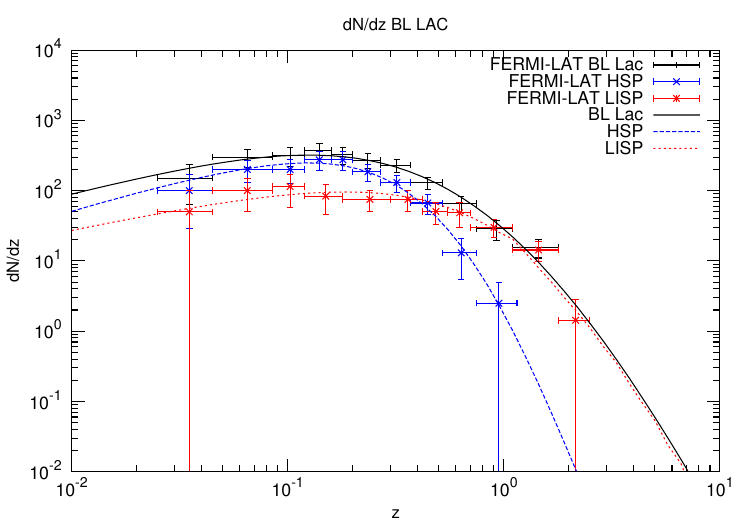}
\includegraphics[width=0.63\textwidth]{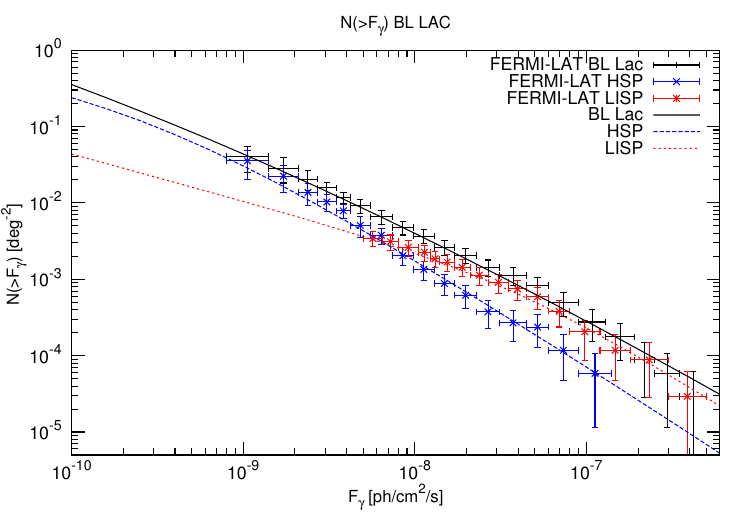}
\caption{From top to bottom: theoretical and experimental luminosity, redshift
and source-count distributions are shown for LISPs (red dotted line/points), HSPs
(blue dashed line/points), and for the whole (LISP+HSP) BL Lac sample (black
solid line/points). Predictions are obtained with the LDDE LF model (Eq.
\ref{eq:ldde1}).} 
\label{fig:distr}
\end{figure}

In Table \ref{tab:parbest} the best-fit parameters with $1\sigma$ uncertainties
of LDDE LF are listed for LISPs, HSPs and the whole BL Lac population. In
Figure \ref{fig:distr} the theoretical and experimental redshift, luminosity and
source-count distributions are shown for LISPs, HSPs and for the BL Lac population
as a whole, and assuming the LDDE LF model. The results in Figure \ref{fig:distr}
show that the HSPs have a narrow redshift distribution peaked around 0.1-0.2,
while the LISPs have a broader distribution that extends to $z\approx2$. Also,
the $\gamma$-ray luminosity distribution is narrower for HSP than for LISP, with
the latter class reaching values of $L_{\gamma}$ at least one order of magnitude
higher. The source-count distribution shows that there are more high-flux LISPs
than HSPs, while the opposite trend is present for low values of $F_{\gamma}$
(for $F_{\gamma}\,=\,10^{-10}$ photons cm$^{-2}$ s$^{-1}$ the number of HSPs is
about a factor of five higher than LISPs). The intensity of the diffuse $\gamma$-ray
emission from unresolved BL Lac relies on the number of sources with a flux
lower than the {\it Fermi}-LAT threshold ($\approx\,10^{-9}$ photons cm$^{-2}$
s$^{-1}$). We expect, therefore, the bulk of the flux from unresolved BL Lacs to
be due to HSPs, which are much more numerous than LISPs for very low
$F_{\gamma}$.

\section{Parameterization of the Spectral energy distributions}\label{SedSec}
\subsection{The Catalog}\label{SedCat}
For the SED parameterization, an implicit hypothesis for any population
study is that some sort of spectral universality (or scaling with other
independent parameters) is present in the sample, or at least in a selected
sub-population. We take as template for this underlying spectrum the average SED
inferred by fitting the single source data to some functional forms. Below, we
describe how these SEDs were obtained. Later on, we shall comment on the
estimated systematic error intrinsic to our theoretical ansatz.

We are interested in obtaining a spectral fit of the three BL Lac sub-populations (LSP,
ISP, HSP). However, one cannot simply take the spectra of the whole sample given
by the 148 objects previously selected, since there is a known ``spectral bias''
to account for: the {\it Fermi}-LAT more easily detects faint sources with a
hard spectrum than with a soft one\footnote{Sources with $\Gamma$=1.5 can
be detected down to fluxes $\gtrsim$20 times fainter than those with a photon
index of 3.0.} \citep{2010ApJ...715..429A}. To avoid this effect, we limit our
spectral analysis to sources with
$F_{100}>F^{\rm{thr}}_{100}\,=\,1.5\times10^{-8}$ photons cm$^{-2}\,{\rm
s}^{-1}$, which has been shown to ensure a detection efficiency independent of
the photon index of the source \citep{2011ApJ...743..171A}.

After this further selection criterion on the catalog selected in
Sect.\ref{lumfun}, we are left with 19 LSP, 21 ISP and 23 HSP BL Lacs, which we
will employ for the averaged SED construction. A total of 57 sources out of the 63
in our sample are present in the First High Energy {\it Fermi}-LAT catalog
\citep[1FHL,][]{2013arXiv1306.6772T} based on 3 years of data. In this case, the
highest energy point of the 2FGL was replaced in the fitting procedure by the
data point of the 1FHL catalog. This ensures that one deals with
statistically independent data points. Spectral information of the
TeV emitters falling in the {\tt default} section of the TeVCat
catalog~\citep{tevcat} \footnote{\url{http://tevcat.uchicago.edu}}
has also been added.

As expected, not all the objects detected at GeV scale have a TeV counterpart: on the
contrary, most of the BL Lac objects in the current TeVCat are HSPs (41), while
we found only 4 LSPs (with 3 FSRQs) and 7 ISPs. This confirms that the HSPs
(sources with the highest value of the synchrotron-peak frequency) also have a
high-energy component peaking from tens of GeV up to TeV energies
\citep{2010ApJ...716...30A,2012arXiv1205.0539B}.
 Note also that among the 41 objects of the TeV catalog labeled as HSPs,
12 fall in the {\tt newly announced} category and their spectra are not publicly
available. Among the remaining 29 objects, 7 are not present in the 2FGL catalog
and are thus excluded in our study. Hence, our final sample of HSPs used to
build the SED consists of 23 sources. 
It is important to note that, especially for the HSPs, the addition of the 1FHL and TeVCat
catalogs---which are not used in the LF determination of
Section \ref{lumfun}---allows us to extend the energy range for the SED
parameterization to the very high energy range.

Finally, we remind the reader that BL Lac objects are variable at all wavelengths. 
While flaring episodes in VHE were detected in the past (e.g., for Mrk 421 or PKS~2155-304), most of the sources do not show variability. 
This may be due to the fact that they were detected close to the sensitivity threshold with deep observations (several tens of hours) spread on few years. 
With the aim to reduce the possible bias introduced by the variability, the lowest flux reported in the literature has been used based on \citet{2009ApJ...707.1310A}, 
where the high and low flux states for flaring sources are reported. 
A similar procedure has been followed in~\citep{2013AA...554A..75S}. A posteriori, a cross-check that this provides a reasonable approach is 
given by the goodness of fits with a single smooth function covering both the averaged/quiet-state Fermi spectrum at low energy and the spectrum
at high energy from Cerenkov telescopes.

\subsection{Spectral Energy Distribution Fits}
\label{SedFit}

\begin{figure}
\centering
\includegraphics[width=0.67\textwidth]{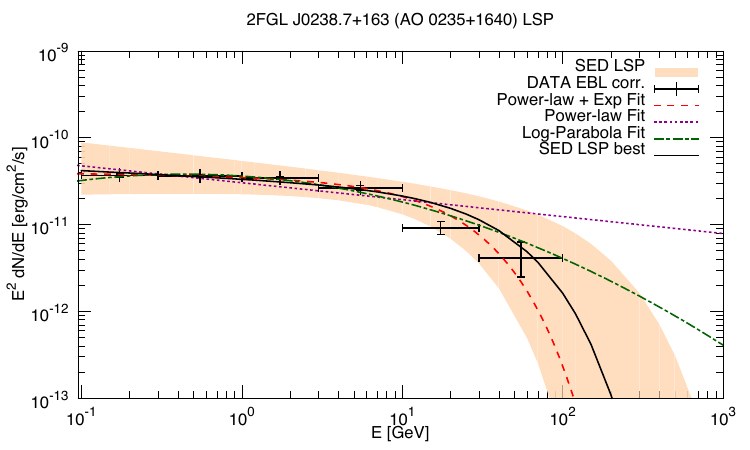}\\
\includegraphics[width=0.67\textwidth]{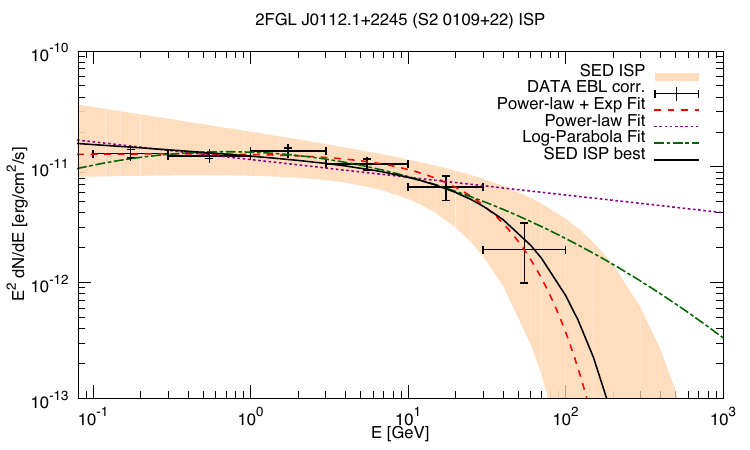}\\
\includegraphics[width=0.67\textwidth]{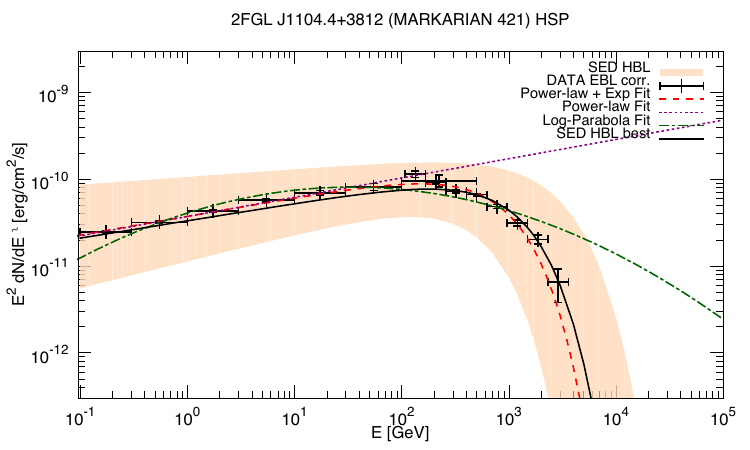}
\caption{Top panel:  SED of 2FGL~J0238, an LSP in our sample.
 The points and curves represent the intrinsic derived spectrum (black points) and the 
 theoretical SEDs computed from fitting the PWL (dotted purple line), LP (dot-dashed green line), and PLEC (dashed red line) functions with the intrinsic SED. 
 We also display the best-fit and the theoretical band (solid black line and pink band) of the SED derived for the LSP sample considering the PLEC model. Middle (bottom) panel: same as top panel,
 for 2FGL~J0112 (Mrk 421), an ISP (HSP) in our sample.}
\label{fig:sedplots}
\end{figure}

The simplest model that can be tested to fit the SEDs is a powerlaw (PWL):
\begin{eqnarray}
\label{eq:powlaw}
   \frac{dN}{dE}\,=\,K\,\left(\frac{E}{E_{0}}\right)^{-\Gamma_{\rm{pow}}},
\end{eqnarray}
where $K$ is a normalization factor, $E_0$ is an arbitrary normalization energy,
and $\Gamma_{\rm{pow}}$ is the photon spectral index. In the 2FGL catalog the
spectra of AGNs are fitted with a log-normal parabola which is more general than
the PWL. This shape is valid for bright sources (mainly AGNs) which are
not very well represented by the powerlaw spectra. The log-normal (LP)
function is given by:
\begin{eqnarray}
\label{eq:lognorm}
   \frac{dN}{dE}\,=\,K'\,\left(\frac{E}{E_{0}}\right)^{-\alpha-\beta\log{\left(\frac{E}{E_0}\right)}},
\end{eqnarray}
where $\alpha$ is the spectral slope at $E_0$ and $\beta$ is the curvature of the spectra.
Finally, we also fit the data with a powerlaw with an exponential cutoff (PLEC):
\begin{eqnarray}
\label{eq:expcut}
   \frac{dN}{dE}\,=\,K''\,\left(\frac{E}{E_{0}}\right)^{-\Gamma_{\rm cut}}\exp{\left(-\frac{E}{E_{\rm cut}}\right)},
\end{eqnarray}
where $E_{\rm cut}$ is the exponential cutoff  energy, and $\Gamma_{\rm cut}$ the powerlaw index associated with this spectral shape.

For HSPs, the data in the combined database (including 1FHL and TeVCat catalogs)
extend up to the very high energy (VHE) range ($E\gtrsim 100\,$GeV). At such
energies, $\gamma$-rays from extragalactic sources have non-negligible
probability to interact with extragalactic background light (EBL) photons. The
intrinsic source spectrum $({dN/dE})_{\rm{intr}}$ is therefore modified because
of the absorption of VHE $\gamma$-rays and the observed spectrum
$({dN/dE})_{\rm{obs}}$ reads
\begin{eqnarray} \label{eq:abs} 
\left(
{\frac{dN}{dE}}\right)_{\rm{obs}}\,=\,\left({\frac{dN}{dE}}\right)_{\rm{intr}}\exp{\left(-\tau_{\gamma\gamma}(E,z)\right)},
\end{eqnarray} 
where $\tau_{\gamma\gamma}(E,z)$ is the optical depth of the EBL.
We adopt the model of \cite{2010ApJ...712..238F} for the EBL to correct the
observed flux. This model is compatible with the recent EBL measurements provided by the Fermi and HESS Collaborations \citep{2012Sci...338.1190A,2013A&A...550A...4H}. We checked also for a sample of relevant redshifts that using the
opacity of~\citet{2008A&A...487..837F} or ~\citet{2012MNRAS.422.3189G}
would have a negligible impact on our results.

\begin{table}
\begin{center}
\begin{tabular}  {||c||c|c||c|c|c||c|c|c||}
\hline
\multicolumn{1}{||c||}{} &
\multicolumn{2}{|c||}{PWL} &
\multicolumn{3}{|c||}{LP} &
\multicolumn{3}{|c||}{PLEC} \\
\hline
&	$\Gamma_{\rm{pow}}$ & $\tilde{\chi}^2$ (d.o.f.) &	$\alpha$ &		$\beta$ &	$\tilde{\chi}^2$ (d.o.f.) &	$\Gamma_{\rm{cut}}$ 	&	$E_{\rm{cut}}$ [GeV]	 & $\tilde{\chi}^2$ (d.o.f.) \\
\hline
\hline
  LSP	&	$2.18 \pm 0.12$	& 3.4	 (114) & $2.13\pm0.13$  &$-1.05\pm0.30$	& 1.25  (113) &	 $2.08 \pm 0.13$ 			&	$34^{+85}_{-20}$&	0.80 (113) \\
  ISP		&	$2.15 \pm 0.14$	&2.6	(130) & $2.12\pm0.14$ &	$-1.08\pm0.33$ & 1.10 (129) &	 $2.07 \pm 0.14$ 			&	$39^{+80}_{-20}$&	0.49 (129) \\
  LISP	&	$2.17 \pm 0.15$	& 3.0 (246) & $2.13\pm0.15$ &$-1.06\pm0.33$	& 1.15 (245)	& $2.08 \pm 0.15$ 			&	$37^{+85}_{-20}$&	0.63 (245) \\
  HSP	&	$1.89 \pm 0.15$	& 5.4	 (248) & $1.87\pm0.14$ &	$-1.25\pm 0.26$ & 1.34 (247) &	 $1.86 \pm 0.16$ 			&	$910^{+1100}_{-450}$ & 0.48 (247)	\\
\hline
\hline
\end{tabular}
\caption{List of the best-fit values and uncertainties of the parameters for the SED of LSP, ISP, LISP, and HSP populations: 
the photon index $\Gamma_{\rm{pow}}$ of PWL, the $\alpha$ and $\beta$ parameters of the LP function, and $\Gamma_{\rm{cut}}$ and  $E_{\rm{cut}}$ for PLEC. 
The corresponding values of the reduced chi-squared ($\tilde{\chi}^2$) and number of d.o.f. are also reported.}
\label{tab:valori}
\end{center}
\end{table}

We fit the intrinsic spectra of our sample of HSPs, ISPs and LSPs with the
PWL, LP and PLEC models. The best-fit values and the $1\,\sigma$ uncertainties
on the best-fit parameters (apart from the normalization ones) for the SED of
LSPs, ISPs, HSPs and for the entire BL Lac population are listed in
Table \ref{tab:valori}. The corresponding values of the reduced chi-squared
($\tilde{\chi}^2$) and number of degrees of freedom are also reported. Note that
the resulting values for the fitted photon index $\Gamma_{\rm pow}$ (PWL) are
very close to the ones found in~\citet{2011ApJ...743..171A}, which reports
$2.17\pm0.12$, $2.13\pm0.14$, and $1.90\pm0.17$ for LSP, ISP and HSP
respectively\footnote{No exact matching is expected, since our data set is
enlarged compared with the one used in~\citet{2011ApJ...743..171A}.}. On the
other hand, using the PLEC model, the photon index $\Gamma_{\rm cut}$ is
systematically harder than in the powerlaw fit case, especially for LSP and
ISP. The HSPs also show systematically harder spectra (by more than 0.2 in the
spectral index) and an inferred cutoff energy that is more than one order of
magnitude higher than for either LSP and ISP objects.

For illustration, in the three panels of Figure \ref{fig:sedplots} we show the SED
for one representative source of each sub-class: an LSP object (top), an ISP
object (middle), and an HSP object (bottom). We display the inferred intrinsic (EBL corrected) spectra (black points),
together with the best-fit PWL, LP and PLEC models for the intrinsic SED.
Averaged best-fit and the 1$\sigma$ error band SED derived for each sub-class
are also presented for comparison in the PLEC model. 

Note that LSPs and ISPs have compatible values for the photon indices $\Gamma_{\rm
pow},\,\Gamma_{\rm cut}$ and the energy of the cutoff $E_{\rm{cut}}$. This is
easily seen from the best-fit results, but can be quantified, for example, by
performing a Student's $t$-test, which yields a compatibility of the two
distributions of about 60\%. We have therefore decided to merge the LSP and ISP
populations into a unique class called LISP, with the following values (shown
also in Table \ref{tab:valori}) for the photon index and energy cutoff:
$\Gamma_{\rm{pow}} = 2.17 \pm 0.15$, $\Gamma_{\rm{cut}} = 2.08 \pm 0.15$ and
$E_{\rm{cut}} =37^{+85}_{-20}$ GeV. Additionally, we checked that the LISP and HSP
populations are not compatible: a Student's $t$-test based on the powerlaw plus
cutoff inferred parameter values yields a compatibility level well below 1\%.

The chi-squared values for the different cases analyzed also show that a PLEC
function provides the best description of the spectra of BL Lacs, among the ones
tested (albeit other models like the PLE can still provide statistically acceptable fits). In general, we find that in several cases a mere powerlaw is not a good
description of VHE data. Note also that for LSP and ISP BL Lacs the average
value of the photon index considering a simple powerlaw is softer than the case
of a powerlaw with an exponential cutoff, confirming the presence of a turnoff
in their spectra. On the contrary, in the case of HSPs the photon index values in
the two cases are not so different, confirming that the cutoff occurs beyond the
energy range explored by the LAT\footnote{For a small number of sources
(for HSP $\approx 20\%$), no statistically significant detection of a cutoff can
be actually claimed, see also the discussion in Section \ref{concl}}.

\begin{figure}
\centering
\includegraphics[width=0.67\textwidth]{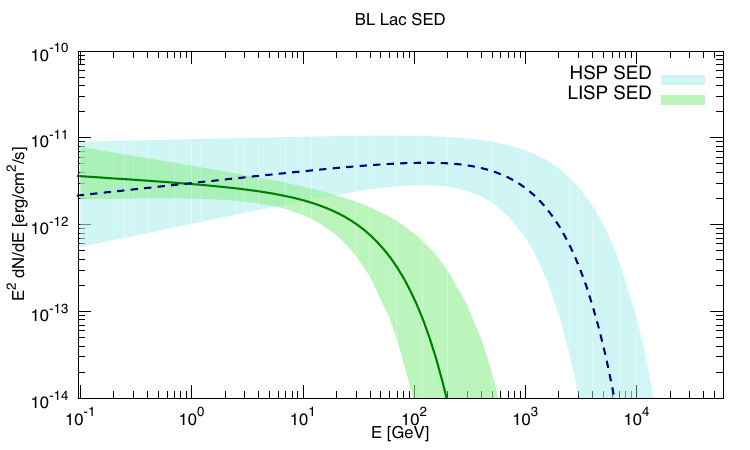}
\caption{best-fit and uncertainty band obtained with the PLEC of the LISP (green) and HSP  (cyan) BL Lac classes. The two spectra are normalized at $3 \times 10$$^{-12}$  erg cm$^{−1}$  s$^{−1}$ at 1 GeV and they represent the intrinsic emission hence without the EBL absorption.}
\label{fig:sedbllac}
\medskip
\end{figure}

Thus, in the following analysis we consider the BL Lac population split into
two sub-classes: LISP with photon index $\Gamma_{\rm cut} = 2.08 \pm 0.15$ and
energy cutoff $E_{\rm{cut}} = 37^{+85}_{-20}$ GeV, and HSP with photon index
$\Gamma_{\rm cut} = 1.86 \pm 0.16$ and energy cutoff $E_{\rm{cut}} =
910^{+1100}_{-450}$ GeV. We show in Fig. \ref{fig:sedbllac} the best-fit and the
uncertainty band for LISP and HSP BL Lac SEDs. The two spectra are normalized to
$3\times 10^{-12}$ erg cm$^{-2}$ s$^{-1}$ at 1 GeV. The uncertainty on the
fitted SED is variable with energy. In particular, for the HSP sample it
is about one order of magnitude at 100 MeV, a factor of three at 100 GeV and it
increases again at higher energies. For comparison, if we were to consider the
BL Lac population globally with spectra given by a powerlaw, we would obtain
$\Gamma = 2.09 \pm 0.20$. Needless to say, a study based on such a simplifying
assumption could provide a too crude representation of the data, especially in
the highest energy range. Yet, in the following sections we shall see the
usefulness of this model to gauge the systematic theory error on our
predictions.

\medskip
\section{Diffuse $\gamma$-ray emission from unresolved BL Lacs}~\label{diffuse}
Armed with the LF and SED previously derived, we are now ready to evaluate the $\gamma$-ray emission arising from faint BL Lacs. 
The contribution of unresolved BL Lacs to the IGRB can be estimated as:
\begin{equation}
\label{eq:diff}
   \displaystyle \Phi_{\rm IGRB}(E_{\gamma}) = \int^{\Gamma_{{\rm max}}}_{\Gamma_{{\rm min}}} 
   d\Gamma \int^{z_{{\rm max}}}_{z_{{\rm min}}} dz \int^{L^{{\rm max}}_{\gamma}}_{L^{{\rm min}}_{\gamma}} dL_{\gamma} \,\Theta_{\gamma}(z,\Gamma,L_{\gamma}) \frac{d F_{\gamma}}{dE} e^{-\tau_{\gamma\gamma}(E,z)} [1-\omega(F_{\gamma})],
\end{equation}
where the limits of integrations are the same as reported in Section \ref{lumfunFit}.
$d F_{\gamma}/dE$ is the intrinsic photon flux at energy $E$, for a BL Lac with $\gamma$-ray luminosity $L_{\gamma}$ \citep{venters2009,yan2012}:
    \begin{eqnarray} \label{Fdef}
          \frac{d F_{\gamma}}{dE} &=& \frac{L_{\gamma}\,(1+z)^{2-\Gamma}}{4\pi {d_L(z)}^2\,E^2_0} \, \mathcal{K},
    \end{eqnarray} 
where $\mathcal{K}$ is a dimensionless spectral normalization factor (calculable numerically) which, for HSP and in the limit $E_{\rm{cut}}\gg E_2,E_1$,  is given by 
\begin{equation} \label{K}
{\cal K}\underset{E_{\rm{cut}}\gg E_2,E_1}{\longmapsto} \frac{(2-\Gamma)}{\left[\left( \frac{E_2}{E_0} \right)^{2-\Gamma}-\left(\frac{E_1}{E_0} \right)^{2-\Gamma} \right]}\,.
    \end{equation}    
In the above formulae, $E_2 = 100$ GeV, $E_1 = 100$ MeV. The LAT detection
efficiency function $(1-\omega(F_{\gamma}))$ \citep{DiMauro:2013xta} enters
in Eq.~\ref{eq:diff} since we are interested in the diffuse flux not resolved by
{\it Fermi}-LAT. Setting $\omega(F_{\gamma}) = 0$ allows us to compute the total
$\gamma$-ray emission arising from the whole BL Lac class, either resolved or
unresolved.

We use two different approaches to predict the flux from unresolved BL Lacs. Our
best estimate is obtained considering the population as made by two
sub-components, the LISP and the HSP BL Lacs, with the SEDs given by PLEC
models. The second method assumes the BL Lac as a unique class of sources with an
SED given by a simple powerlaw. This (overly)simplistic model should provide an
estimate of possible ``systematic'' biases coming from current misunderstanding
of the BL Lac population, in particular in what concerns extrapolations to low
fluxes or luminosities. We also calculate the uncertainty band for each case
considering the 1$\sigma$ errors on the LF and SED parameters for LISPs, HSPs
and BL Lacs given in Tabs.~\ref{tab:valori} and \ref{tab:parbest}.

\begin{figure}
\centering
\includegraphics[width=0.85\textwidth]{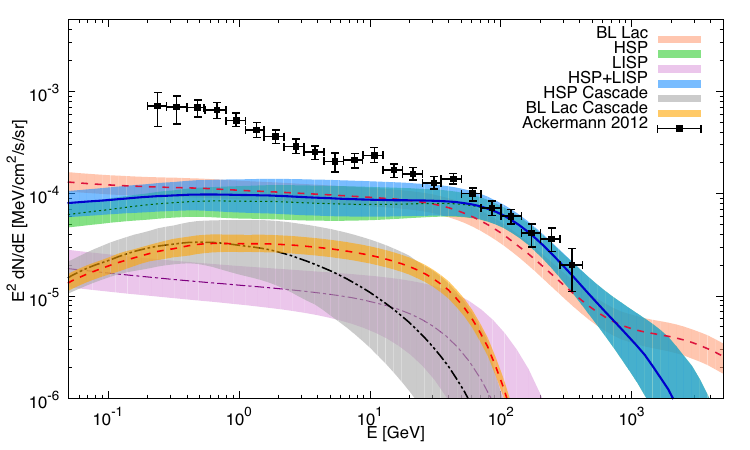}
\caption{Diffuse $\gamma$-ray emission from unresolved BL Lacs. Predictions for the best-fit models (see text for details) are shown embedded in their 1$\sigma$ uncertainty bands: 
the contribution is displayed for LISP (purple dot-dashed curve), HSP  (dotted green), the sum of LISP and HSP (solid blue) and for BL Lacs considered as a unique population (dashed red). 
The IGRB data \citep{ackermann2012} are also displayed with black points. The black dot-dashed (dashed yellow) line represents the cascade emission from the HSP (the whole 
BL Lac population) source flux.}
\label{fig:lffit}
\end{figure}

Our main results are reported in Figure \ref{fig:lffit}, along with the {\it
Fermi}-LAT IGRB data~\citep{ackermann2012}. We present the predictions for
the diffuse $\gamma$-ray emission from unresolved BL Lacs both when treated as a
unique population and considering the two sub-populations of LISP and
HSP separately. For the latter case, the separate LISP and HSP contributions are
displayed together with their sum. The high-energy LISP and HSP flux is shaped
by the cutoff of the SEDs, while for the case of the BL Lacs treated as a unique
population only the $\gamma$-ray EBL absorption intervenes in softening the
spectrum (their SED being described by a simple powerlaw) for energies $\gsim$
30 GeV. Overall, the excellent agreement between our fiducial prediction
(LISP+HSP) and the data suggests that unresolved HSP BL Lacs account for the
largest fraction of the IGRB beyond about 50 GeV. Remarkably,
Figure \ref{fig:lffit} indicates the even the simplistic/extreme model with a
single population and powerlaw extrapolated SED would give a similar (albeit
not as good) result in the energy decade around 100 GeV, although it
overestimates the flux for the lowest energies and underestimates the flux at
high energies. The EBL absorption alone can justify the measured decreasing
trend in the IGRB spectrum above 100 GeV, albeit the shape suggested by the HSP
SED (harder spectrum with exponential cutoff, see Section \ref{SedFit}) seems to
better reproduce the detailed trend. 

The (very) high-energy radiation absorbed on the EBL via pair production may trigger a cascade photon production. 
The  resulting $e^{+}e^{-}$ pairs, depending on the strength of the environmental
magnetic fields, lose energy either via synchrotron radiation or via the inverse Compton process onto the cosmic microwave background (CMB). In the latter
case, the up-scattered CMB photons end up typically at GeV energies and can contribute to the IGRB. 
We estimated this contribution using the formalism described e.g. in~\citet{Inoue:2012cs} and applied to the mis-aligned AGN diffuse emission in 
\cite{DiMauro:2013xta}, assuming that all the absorbed energy at VHE ends up in those re-scattered photons. We also 
set to zero the intergalactic magnetic field, which implies an upper limit for the cascade emission at the $\gamma$-ray energies. 
The results for the cascade photons are shown in Figure \ref{fig:lffit} with a gray (ochre) band for the 
HSP (BL Lac considered as a unique) population. This contribution amounts to no more than $\sim 30\%$ of the primary flux, and 
 additionally in the GeV range where the BL Lac contribution to the 
IGRB is anyway subdominant. It is worth nothing that the amount of this ``tertiary'' 
radiation depends on the amount of energy absorbed at VHE and hence on the hardness of the primary spectrum 
(the larger background expected at low energies for the baseline model follows from the HSP hard spectrum) and
on the presence (and value) of an energy cutoff. 
Also note the larger dispersion of the prediction (especially at high energy) for the HSP case, due to the strong impact of $E_{\rm cut}$. 
Despite its marginal role, the cascade contribution has been included in all the fluxes displayed in Fig. ~\ref{fig:lffit}.

\bigskip
\section{Discussion and conclusion}
\label{concl}
%

In this paper, we present a new calculation of the diffuse $\gamma$-ray
emission expected from the unresolved BL Lac blazars. For the first time, we
have focused the attention on the ${\cal O}(100)$ GeV energy range, pushing the
predictions up to the TeV scale. To that purpose, while most of the statistical
population properties rely on {\it Fermi}-LAT data, the synergy with IACT data
has proven essential in modeling the spectrum of these objects at the energies
of interest. The IACT data are fundamental in two respects: a)
in confirming that photons are indeed injected from BL Lacs up to the TeV
energy, thus removing the {\it ansatz} of high-energy extrapolation; b) in
refining the spectral shape determination. The HSP+LISP prediction in
Figure \ref{fig:lffit} is clearly in a good agreement with available data, within
the estimated uncertainty band. Notably, our best estimation seems to {\it
fully} account for the measured IGRB at $E_\gamma\gtrsim50\,$GeV: note that
there is no free-parameter adjustment of prediction to data in
Figure \ref{fig:lffit}! Our detailed analysis thus confirms quantitatively the conjecture put forward in some earlier works as ~\citep{Neronov:2011kg}.
Also, the role of BL Lacs is relatively sub-leading at low
energies: for our fiducial best-fit model, the estimated contribution to the
measured IGRB between 100 MeV and 100 GeV amounts merely to $\sim$11\% of the
{\it Fermi}-LAT data \citep[in agreement with the results of][]{Ajello:2013lka}.

Additionally, our analysis suggests that for (very) high energy $\gamma$-ray
purposes it is meaningful to distinguish between HSP and the joint LSP and ISP
(LISP) sub-classes. Considering HSP and LISP as a unique population would still
lead to an acceptable prediction for the diffuse emission in the energy band
considered, but at the expense of a relatively worse SED fit (see
Table \ref{tab:valori}). The resulting prediction also falls a bit short of the
measured IGRB, albeit the statistical significance of the difference is not
high, yet. Also, the extrapolations to lower and higher energies are
significantly different, potentially with implications for other aspects of
$\gamma$-ray astrophysics. 

Concerning the model uncertainties, we included in our predictions the
1$\sigma$ errors coming from the LF and SED fits. A handle on the
``systematic'' theoretical uncertainties is notoriously more difficult to
achieve. Yet, our exercise of using the oversimplified single population and
powerlaw SED model leads to comparable results, strengthening our confidence in
the predictions. Most likely, at least in the one decade around 100 GeV, the
systematic errors are not larger than the statistical ones. Other causes of
error are expected to be sub-leading: the "completeness function'' $\omega$ is
itself derived from data and is subject to some uncertainty. However, its impact
on the unresolved flux is very modest (we estimate $\sim {\cal O} (5\%))$, since
the factor $(1-\omega)$ entering in Eq.~\ref{eq:diff} is very close to 1 at the
low luminosities that contribute the most to the unresolved flux. Similarly,
current observations by {\it Fermi}-LAT~\citep{Ackermann:2012sza} or
HESS~\citep{2013A&A...550A...4H} corner quite well the range of
opacities of the Universe, at least at relatively low redshifts of major
interest for our purpose. At TeV energies or beyond, however, more extreme
assumptions on the level of EBL that may be still marginally allowed by the
data could nonetheless affect our predictions by an amount comparable to the
size of the error band shown in Figure \ref{fig:lffit}. A more extensive study of
those uncertainties may be justified, once IGRB data in that energy region will
become available, and is left for future work.

\begin{figure}
\centering
\includegraphics[width=0.85\textwidth]{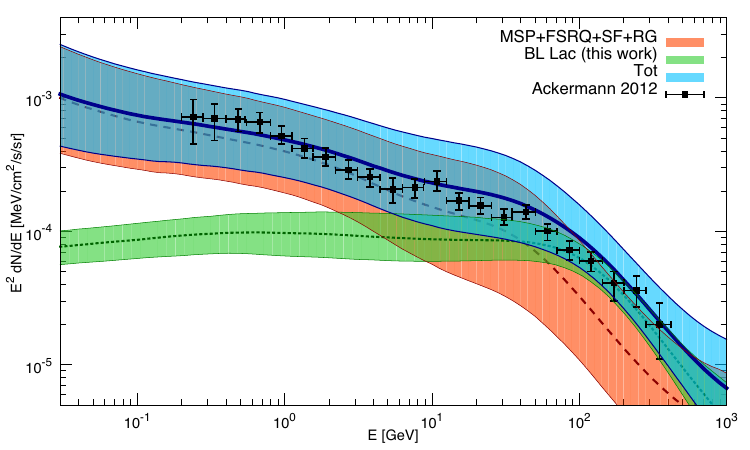}
\caption{Global view of the diffuse $\gamma$-ray predictions (best-fit models
and relevant uncertainty band) is displayed for unresolved BL Lac
(dotted green, this work) and for the sum of misaligned AGN \citep[]{DiMauro:2013xta}, star forming galaxies \citep[]{2012ApJ...755..164A}, FSRQs 
\citep[]{2012ApJ...751..108A}, and millisecond pulsars 
\citep[]{Calore:2011bt} (orange dashed line and
uncertainty band). IGRB data \citep{ackermann2012} are also
displayed with black points. The sum of all the predictions is displayed in a blue
curve line and cyan uncertainty band. }
\label{fig:all}
\end{figure}

Globally, the emerging picture seems to be that the IGRB may be naturally
explained by the cumulative $\gamma$-ray emission from unresolved extragalactic
(and, to some extent, galactic) sources, as already discussed in the
Introduction. Needless to say, estimates exist in the literature for the
contributions to the IGRB of other populations of unresolved astrophysical
sources. Among them, FSRQs \citep{2012ApJ...751..108A}, misaligned AGNs
\citep{2011ApJ...733...66I,DiMauro:2013xta}, star forming galaxies
\citep{2012ApJ...755..164A} and galactic millisecond pulsars
\citep{Calore:2011bt}. Without the aim of completeness, it can be instructive to
show in Figure \ref{fig:all} the unresolved $\gamma$-ray emission from the above-cited populations together with the one for BL Lacs computed in the present work.
It is remarkable that the predicted best estimate (which is the sum of the best
model for each population) is extremely close to the measured data points, which
are found within the estimated uncertainty. This agreement extends over more
than three decades of energy, suggesting that there is relatively little
room for major extra sources of this diffuse background. The importance of strengthening these
predictions in the future, via a consolidation a multi-messenger GeV-TeV
cosmology thanks to instruments such as CTA \citep[see for
example][]{2013APh....43..103R,2013APh....43..317D}, should not be
underestimated. In fact, apart from being a ``sanity check'' of our current
understanding of {\it populations} of astrophysical accelerators, the IGRB has
already been suggested to act as a diagnostic tool for subfields of
astroparticle physics as diverse as ultra-high-energy cosmic-rays
\citep[se e.g.][]{Kalashev:2007sn} or indirect dark matter searches
\citep[as recently in][]{Calore:2011bt,2013arXiv1303.3284B}. 

As an intriguing example, we note that  for a minority (roughly 20\%) of
the HSP sources in our sample, the fitting procedure did not find any
indication of cutoff in the spectra. This feature could be just an
artifact of the limited sensitivity at high energies, or might hint to
the existence of a new sub-class of sources (sometimes dubbed  ultra-high-frequency-peaked BL Lac~\citep{1999ApL&C..39...17G,Senturk:2013pa}). In such a case, some departure from the baseline flux prediction presented in Figs. \ref{fig:lffit}-\ref{fig:all} could most easily show up at TeV energies.

In this context, it is worth commenting on the fact that in recent years
several authors have studied an alternative mechanism for generating hard TeV
spectra from distant BL Lacs: line-of-sight interactions of cosmic rays
("protons") with cosmic microwave background radiation and EBL can
generate secondary gamma rays relatively close to the observer and with
hard spectra (see e.g. Refs.~\citep{Essey:2009zg,Murase:2011cy,Prosekin:2012ne,Zheng:2013lza}).
This is certainly an example of a scenario in which both the SED at VHE
(even harder spectra, much higher cutoff energy...) and the LF (e.g.
relatively large detectable population at large z) could depart from the
minimal, two main population scenario discussed above. Interestingly, this would
also open up a different diagnostic tool of gamma {\it propagation},
since cascades are much more sensitive to ``environmental'' parameters
such as the EBL and the extra-galactic magnetic fields.  Fortunately, it
has been argued that future surveys have the potential to uncover these
populations (see fe.g. ~\citep{Inoue:2013vpa}). No doubt, the next decade in VHE
gamma astrophysics may still reserve some surprises!

\begin{acknowledgments}
We thank M. Ajello and L. Latronico  for useful discussions.  
At LAPTh and LAPP, this activity was supported by the Labex grant ENIGMASS. F.D. further acknowledges support from AAP-``LeSAdHE'' of Univiversity of Savoie. 
We thank Arman Esmaili and Marco Regis for spotting typos in the Appendix.
\end{acknowledgments}

\appendix
\section{Model parameterization tested for the LF fits}\label{appA}
We have tested several models to fit the LF, notably the pure luminosity evolution (PLE), the luminosity-dependent density evolution (LDDE) and 
the steep-spectrum radio sources models (SSRS). For the sake of completeness, we report here their parameterizations. 

Our best-fit is obtained for the LDDE model~\citep{2003ApJ...598..886U},  in which the evolution is primarily in density, with a luminosity-dependent redshift peak.  Its parameterization is:
\begin{eqnarray}
\label{eq:ldde1}
   \rho(z,L_{\gamma})= \rho(L_{\gamma}) \, e(z,L_{\gamma}) ,
\end{eqnarray}
with 
\begin{eqnarray}
\label{eq:ldde0}
   \rho(L_{\gamma})= 
	\frac{A}{\log{(10)}L_{\gamma}}\left[ \left( \frac{L_{\gamma}}{L_{c}} \right)^{\gamma_1} + \left( \frac{L_{\gamma}}{L_{c}} \right)^{\gamma_2} \right]^{-1} ,
\label{eq:ldde2}\\
   e(z,L_{\gamma}) = \left[ \left( \frac{1+z}{1+z_c(L_{\gamma})} \right)^{p_1} + \left( \frac{1+z}{1+z_c(L_{\gamma})} \right)^{p_2} \right] ,
\end{eqnarray}
and
\begin{eqnarray}
\label{eq:ldde3}
  z_c(L_{\gamma}) = z^{\star}_c\, \left( \frac{L_{\gamma}}{10^{48} \rm{\,erg\,s}^{-1}}\right)^{\alpha} \,.
\end{eqnarray}

The best-fit parameters and  $1\sigma$ uncertainties for this model are reported in Table~\ref{tab:parbest}.
\begin{table*}
\begin{center}
\begin{tabular}  {|c|c|c|c|c|c|c|c|c|c|}
\hline
 		&	$A\,(10^{-9} {\rm Mpc}^{-3} {\rm erg}^{-1} {\rm s})$		& $L_c ( 10^{45} {\rm erg/s})$	 & 	$\gamma_1$		& 	$\gamma_2$		& 	$p_1$			&	$p_2$			&	$z^{\star}_c$		&	$\alpha$	 & $\tilde{\chi}^2$ (d.o.f.)	\\
\hline
\hline
  LISP	&	$4.37 \pm 0.78$	& $30.8 \pm 5.7$ 	&  $1.19^{+0.26}_{-0.16}$	& $0.67^{+0.11}_{-0.16} $ &  $4.4^{+2.1}_{-1.1}$	&	$-2.9^{+1.3}_{-1.0}$	&$1.66^{+0.64}_{-0.49}$
		&	$0.36^{+0.15}_{-0.19}$   & 0.37 (32)\\
\hline
\hline
  HSP	&	$98 \pm 18$	 & $3.15 \pm 0.45$  & $2.88^{+1.32}_{-0.75}$	& $0.52^{+0.14}_{-0.17}$ & $-1.64^{+0.59}_{-0.56}$	&	$4.8^{+1.4}_{-2.0}$	&	$4.1^{+1.6}_{-2.2}$
		&	$0.25^{+0.11}_{-0.05}$  & 0.34 (30)\\
\hline
\hline
  BL Lac	&	$96 \pm 13$	 & $1.82 \pm 0.22$ & $0.59^{+0.15}_{-0.21}$ &	$1.43^{+0.11}_{-0.08}$ &	$1.54^{+0.86}_{-0.64}$ &	$-0.42^{+0.23}_{-0.25}$ & $2.10^{+1.20}_{-1.25}$
		&	$0.052^{+0.030}_{-0.022}$  & 0.49 (39)\\
\hline
\hline
\end{tabular}
\caption{The best-fit parameters with $1\sigma$ uncertainties for LDDE LF are listed for LISP, HSP and BL Lac together with the value of the reduced chi-square $\tilde{\chi}^2$}.
\label{tab:parbest}
\end{center}
\end{table*}

The PLE model \citep{2003ApJ...598..886U,2005A&A...441..417H} is given by the following parameterization:
\begin{eqnarray}
\label{eq:ple1}
   \rho(z,L_{\gamma})= \rho(L_{\gamma}/e(z)) ,
\end{eqnarray}
where:
\begin{eqnarray}
\label{eq:ple2}
   \rho(L_{\gamma}/e(z=0))= \frac{dN}{dL_{\gamma}} = 
	\frac{A}{\log{(10)}L_{\gamma}}\left[ \left( \frac{L_{\gamma}}{L_{c}} \right)^{\gamma_1} + \left( \frac{L_{\gamma}}{L_{c}} \right)^{\gamma_2} \right]^{-1} ,
\end{eqnarray}
and
\begin{eqnarray}
\label{eq:ple3}
   e(z) = (1+z)^k e^{z/\xi}.
\end{eqnarray}
In this model, the evolution is entirely in luminosity, thus the sources were more luminous in the past if a positive evolution ($k> 0$) is found (the opposite is true otherwise). 
It is also straightforward to demonstrate that the luminosity evolution peaks at $z_c =-1 -k\xi$. 

Finally we checked a the  SSRS model \citep{2001MNRAS.322..536W}, where the LF is divided into two components:
\begin{eqnarray}
\label{eq:rg1}
   \rho(z,L_{\gamma})= (\rho(L_{\gamma})_I + \rho(L_{\gamma})_{II}) \, f(z) ,
\end{eqnarray}
where:
\begin{eqnarray}
\label{eq:rg2}
   \rho(z,L_{\gamma})_I = \rho_{I}^{0} \left(\frac{L_{\gamma}}{L_{Ic}}\right)^{-\alpha_I} \exp{\left(-\frac{L_{\gamma}}{L_{Ic}}\right)},
\end{eqnarray}
\begin{eqnarray}
\label{eq:rg3}
  \rho(z,L_{\gamma})_{II} = \rho_{II}^{0} \left(\frac{L_{\gamma}}{L_{IIc}}\right)^{-\alpha_{II}} \exp{\left(-\frac{L_{IIc}}{L_{\gamma}}\right)},
\end{eqnarray}
and:
\begin{eqnarray}
\label{eq:rg4}
  	f(z) =
	\left\{
\begin{array}{ll}
 (1+z)^k & \mbox{if } z \leq z_0 \\
 (1+z_0)^k & \mbox{if } z > z_0
\end{array}
\right.
\end{eqnarray}
This model has been used, for instance, 
 by \cite{2011ApJ...733...66I} and \cite{DiMauro:2013xta} for the determination of the diffuse $\gamma$-ray emission from 
  misaligned AGNs.

\section{HSP Catalog for SED determination}\label{appB}
\begin{table*}
\begin{center}
\begin{tabular}  {|c|c|c|c|c|c|c|}
\hline
 Name			&		2FGL name 		&	z 		&	$\Gamma_{\rm HE}$	&	$\Gamma_{\rm VHE}$	&	$\Gamma_{\rm cut}$    &	$E_{\rm cut}$ [GeV] \\
\hline
\hline
RGB J0152+017	&	J0152.6+0148	&	0.080	&	$1.79 \pm 0.14 $		&	$2.95 \pm 0.41$	&	1.82$\pm$0.13		&	1970$^{3300}_{1200}$	\\			
RBS 0413			&	J0319.6+1849	&	0.190	&	$1.55 \pm 0.11 $		&	$3.18 \pm 0.74$	&	1.75$\pm$0.12			&	630$^{540}_{290}$		\\	
1ES 0414+009		&	J0416.8+0105	&	0.287	&	$1.98 \pm 0.16 $		&	$3.45 \pm 0.32$	&	1.83$\pm$0.04			&	//		\\	
PKS 0447-439		&	J0449.4-4350	&$0.25\pm0.10$&	$1.86 \pm 0.02 $	&	$3.89 \pm 0.43$	&	1.88$\pm$0.02		&	370$^{240}_{75}$	\\	
VER J0648+152	&	J0648.9+1516	&	0.179	&	$1.74 \pm 0.11 $		&	$4.4 \pm 0.9$		&	1.94$\pm$0.26			&	450$^{510}_{230}$		\\	
RGB J0710+591	&	J0710.5+5908	&	0.125	&	$1.53 \pm 0.12 $		&	$2.69 \pm 0.33$	&	1.67$\pm$0.07			&	2760$^{2100}_{1100}$	\\	
1ES 0806+524		&	J0809.8+5218	&	0.138	&	$1.94 \pm 0.06 $		&	$3.6 \pm 1.0$		&	1.86$\pm$0.05			&	510$^{230}_{160}$		\\	
1ES 1011+496		&	J1015.1+4925	&	0.212	&	$1.72 \pm 0.04 $		&	$4.0 \pm 0.5$		&	1.84$\pm$0.02		&	355$^{190}_{120}$		\\	
1ES 1101-232		&	J1103.4-2330	&	0.186	&	$1.80 \pm 0.21 $		&	$2.94 \pm 0.20$	&	1.83$\pm$0.25			&	//		\\	
Markarian 421		&	J1104.4+3812	&	0.031	&	$1.77 \pm 0.01 $		&	$2.20 \pm 0.08$	&	1.785$\pm$0.006			&	692$^{34}_{32}$		\\	
Markarian 180		&	J1136.7+7009	&	0.045	&	$1.74 \pm 0.08$		&	$3.25 \pm 0.66$	&	1.75$\pm$0.07			&	905$^{570}_{350}$		\\	
1ES 1215+303		&	J1217.8+3006	&	0.13		&	$2.02 \pm 0.04$		&	$2.96 \pm 0.21$		&	2.15$\pm$0.03			&	2000$^{2200}_{1040}$	\\	
1ES 1218+304		&	J1221.3+3010	&	0.182	&	$1.71 \pm 0.07$		&	$3.08 \pm 0.39$	&	1.78$\pm$0.06			&	1010$^{1040}_{510}$	\\	
H 1426+428		&	J1428.6+4240	&	0.129	&	$1.32 \pm 0.12$		&	$3.5 \pm 0.35$	&	1.05$\pm$0.16			&	570$^{190}_{140}$	\\	
PG 1553+113		&	J1555.7+1111	&	[0.43,0.58]		&	$1.67 \pm 0.02$		&	$4.5 \pm 0.3$		&	1.771$\pm$0.016			&	//		\\	
Markarian 501		&	J1653.9+3945	&	0.034	&	$1.74 \pm 0.03$		&	$2.72 \pm 0.18$	&	1.835$\pm$0.014			&	1630$^{350}_{280}$	\\	
1ES 1959+650		&	J2000.0+6509	&	0.048	&	$1.94 \pm 0.03$		&	$2.58 \pm 0.18$	&	1.98$\pm$0.02			&	2750$^{1600}_{1020}$	\\	
PKS 2005-489		&	J2009.5-4850	&	0.071	&	$1.78 \pm 0.05$		&	$3.20 \pm 0.19$	&	1.77$\pm$0.04			&	220$^{32}_{28}$		\\
PKS 2155-304		&	J2158.8-3013	&	0.116	&	$1.84 \pm 0.02$		&	$3.32 \pm 0.06$	&	1.864$\pm$0.011			&	350$^{30}_{25}$		\\	
B3 2247+381		&	J2250.0+3825	&	0.119	&	$1.84 \pm 0.11$		&	$3.2 \pm 0.7$		&	1.77$\pm$0.13			&	//		\\	
1ES 2344+514		&	J2347.0+5142	&	0.044	&	$1.72 \pm 0.08$		&	$2.95 \pm 0.23$	&	1.71$\pm$0.07			&	630$^{170}_{130}$	\\	
H 2356-309		&	J2359.0-3037	&	0.165	&	$1.89 \pm 0.17$		&	$3.09 \pm 0.26$	&	1.71$\pm$0.11			&	3720$^{4700}_{2100}$	\\	
\hline
\hline
\end{tabular}
\caption{List of the parameters for the sample of 23 HSP BL Lacs selected for the determination of the SED:  name, 2FGL name, redshift, {\it Fermi}-LAT photon index from 100 MeV to 100 GeV $\Gamma_{\rm HE}$, 
VHE photon index measured at TeV energies by Cerenkov experiments $\Gamma_{\rm VHE}$, and photon index $\Gamma_{\rm cut}$ and energy 
cutoff $E_{\rm cut}$, both fitted {\it in our analysis} according to  Eq.~\ref{eq:expcut} .
The values associated to $\Gamma_{\rm HE}$, $\Gamma_{\rm VHE}$ and $z$ are taken from \citep{2013AA...554A..75S} for all sources except 
for 1ES 1215+303 \citep{2012AA...544A.142A}, PKS 0447-439 \citep{1998AJ....115.1253P,2012AA...543A.111P,2013arXiv1307.6907R,2013arXiv1309.2386Z,2013AA...552A.118H}, PG 1553+113 \citep{2010ApJ...720..976D,2008AA...477..481A} and Mrk 501 \citep{2011ApJ...729....2A}. }
\label{tab:hsp}
\end{center}
\end{table*}
For completeness, in Table \ref{tab:hsp} we report the characteristic parameters for the selected HSP BL Lacs, 
which constitute the class of greatest interest for this work, and where the synergy between HE and VHE observations is more evident: 
associated name, 2FGL name, redshift, photon index in the range 100 MeV-100 GeV $\Gamma_{\rm HE}$, 
photon index in the TeV energies $\Gamma_{\rm VHE}$, photon index for $E>100$ MeV $\Gamma_{\rm cut}$ and 
energy cutoff  $E_{\rm cut}$ found with our analysis for  an exponentially cutoff powerlaw  spectrum (see Eq.~\ref{eq:expcut}).
The values associated with $\Gamma_{\rm HE}$, $\Gamma_{\rm VHE}$ and $z$ are taken from \citep{2013AA...554A..75S} and references therein
for all sources except for 1ES 1215+303 \citep{2012AA...544A.142A}, 
PKS 0447-439 \citep{1998AJ....115.1253P,2013arXiv1307.6907R,2013arXiv1309.2386Z,2013AA...552A.118H}, PG 1553+113 \citep{2010ApJ...720..976D,2008AA...477..481A} and Mrk 501 \citep{2011ApJ...729....2A}.

\bigskip

\bibliography{paper}

\end{document}